# Numerical Analysis of Liquid Cooling of 3D-ICs Using Embedded Channels

Sakib Islam[1] and Ibrahim Abdel-Motaleb[2]
*Department of Electrical Engineering, Northern Illinois University , Dekalb, USA*
[1]aislam@niu.edu , [2] ibrahim@niu.edu

*Abstract*— Hot-spots are considered among the unavoidable consequences of the high integration density of 3D-ICs. Eliminating hotspots requires employing efficient cooling techniques. Using embedded channels, liquid cooling systems can be designed to deliver the right amount of coolant to each spot of the chip. In this study, numerical analysis is used to investigate the cooling of a 20 W hotspot using embedded channels employing three coolants: water, Freon (R22), and liquid nitrogen (LN). The investigation of thermal management and stress show that, although LN provides the lowest operating temperature (164 K), it causes the highest stress (355 MPa) at 100 mm/s inlet velocity. The study also shows that, coolant delivery using parallel channels results in a wide variation of local temperatures and stress. This stress variation may form "high-stress spots," which may cause circuit failure, performance degradation, or yield reduction. Therefore, cooling systems and chip fabrication should be designed to ensure the elimination of high-stress hotspots.

## I. Introduction

Increasing the packing density requires the reduction of the device size. Published reports show that device size has reached its physical limit and Moore's law is almost reaching its end. Although several technologies, such as quantum computing or single electron transistor, have been proposed to meet the increasing demand for high integration density, the 3D-IC technology is believed to be the most practical one. 3D-ICs are built by stacking several 2D-ICs on top of each other and connecting them using metal interconnects through vias [1]. Such arrangement increases the integration density by as many folds as the number of the stacked ICs. 3D-IC technology can also provide reduced latency, multi-tasking, high speed processing, and heterogeneous integration [2]. The main problem with 3D-ICs, is the creation of the hot-spots, with local temperatures reaching very high temperatures. These hotspots may result in dysfunctional operation, component failure [3], thermal stress, signal delays or chips' permanent degradation. Although several cooling measures, such as Microelectromechanical Systems (MEMS)-based technology, embedded micro channels [5,6], liquid immersion cooling [7], and microfluidic cooling using thermal Through Silicon Vias (TSVs) [8] have been proposed, they may not be able to eliminate these thermal challenges completely.

In this study, thermal stress analysis is conducted for a chip with 20 W hot-spot that is liquid-cooled through parallel embedded channels. The study was conducted using Comsol, a multi-physics numerical analysis program. Water, Freon R22, and Liquid Nitrogen (LN) are used as coolants.

## II. Heat Transfer and Thermal Stress

The heat transfer in any system can be a result of a combination of one or more of the following mechanisms: conduction, convection, and radiation. Heat conduction through a material with length $l$ and cross-section $A$ can be expressed by the following equation.

$$\frac{Q}{t} = kA\frac{\Delta T}{l} \qquad (1)$$

Here, $Q$ is the transferred heat, $t$ is the time, $k$ is the thermal conductivity (W.m$^{-1}$K$^{-1}$) and $\Delta T$ is the temperature gradient, or the temperature difference between the two ends of the object in Kelvin (K). $Q/t$ is the power in watts.

Convection is the heat flow carried by the motion of fluid or gas. Heat convection can be expressed by the following equation.

$$Q = hA\Delta T \qquad (2)$$

In this equation, $h$ is the convective heat transfer coefficient (J.m$^{-2}$K$^{-1}$), $A$ is the object's cross sectional area which carries the fluid, and $\Delta T$ is the temperature difference between fluid temperature, $T_F$, and heated surface temperature, $T$.

Radiation is a method of heat transfer, where the exchange of heat takes place through heat emission to the ambience. Radiation can be described by the following equation,

$$P = \varepsilon\sigma AT^4 \qquad (3)$$

Here, $P$ is the power, $\varepsilon$ is the emissivity factor, $\sigma$ denotes Stefan-Boltzmann constant (=5.67 X 10$^{-8}$ W.m$^{-2}$.K$^{-4}$), $A$ is the surface area of the radiating body and $T$ is the temperature of the radiating body [9].

The relationship between stress and strain is bound by the following equation,

$$\sigma = \varepsilon E \qquad (4)$$

In this equation, $\sigma$ is the uniaxial stress, $\varepsilon$ is the strain or deformation, and $E$ is Young's modulus, measures the stiffness of a solid material [10,11].

Heat may cause thermal stress. Thermal stress can be expressed by the following equation,

$$\sigma = E\alpha\Delta T = E\alpha(T_F - T_0) \qquad (5)$$

where $\alpha$ is the thermal expansion coefficient, $T_0$ is the initial and $T_F$ is the final temperature of the object, and $\Delta T = (T_F - T_0)$ in kelvins (K). The object temperature, $T_F$ varies with respect to time due to heat transfer, resulting in temperature difference [12].

## III. DESIGN OF THE COOLING SYSTEM

For this study, a 20 W hotspot is represented by a 20 W tungsten heater [13]. Fig. 1 shows the heater and the Cu wire used in this study. The heater is built on a 5x5 mm² Si-substrate with 200 μm thickness. On top of the wafer, a 3000 Å layer of $SiO_2$ is grown, on top of which, a 0.18μm -thick serpentine-shaped tungsten heater and Cu wires are deposited, separated by a distance of 450μm. The area of the rectangle, representing the outer boundaries of the heater and the Cu wire, is about 1000x2000 μm². On top of the metals, 0.18 μm of $SiO_2$ is deposited. Two 0.18μm-aluminum pads with an area of 125x250 μm² are deposited at the ends of the of the heater and connected through the $SiO_2$ layer to the heater vias. Similar pads of Cu were deposited at the end of the Cu wire and connected through vias too. Finally, a 7000 Å $SiO_2$ passivation layer was deposited on the top.

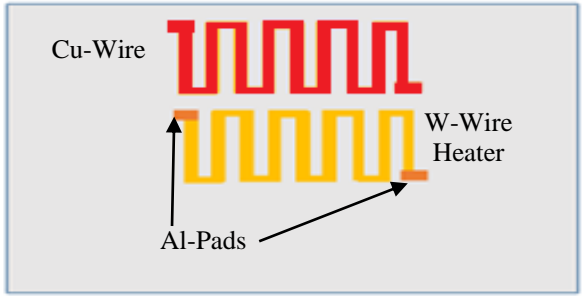

Fig.1: Bottom: Serpentine Shaped Tungsten Heating Block (Yellow); Top: Copper Wire (Red).

The cooling block is made using two bonded Si substrates with a total width of 600 μm and an area of 5x5 mm². One 2D-IC is sandwiched between two blocks, as shown in Fig. 2 (a). To build a 3D-IC, the stacks and the blocks alternate one on top of the other, terminated with a cooling block on top and one at the bottom. The cooling block has embedded channels with the dimensions shown in Fig. 2(b).

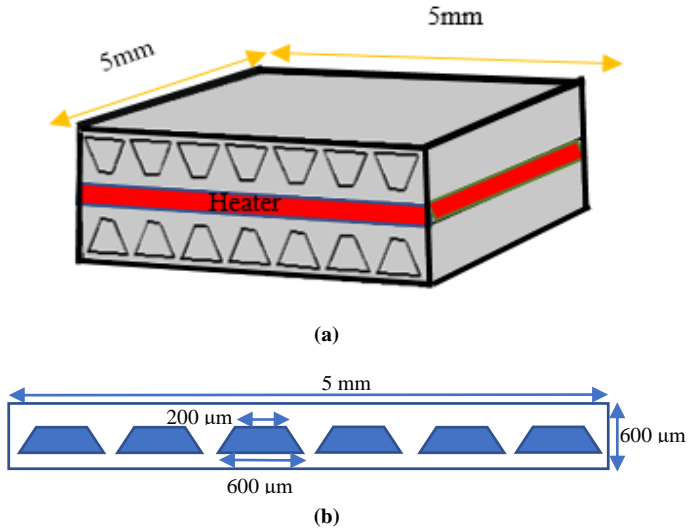

Fig. 2: **(a)** Heater sandwiched between two cooling blocks **(b)** Cooling block geometry; inlets are 100μm above the bottom of the Si Substrate. Inclined lines are 350μm.

## IV. SIMULATION AND ANALYSIS

The thermal analysis was done using the physical material parameters provided by the Comsol material library unless it is otherwise stated. The emissivity value used in the simulation for Si is 0.93 [14]. The convective heat transfer coefficient is set to 50 W/m²K for all the coolants and 10 W/m²K for air. If the heater is turned on without admitting any coolant, the chip temperature would reach 2170K and the thermal stress would reach 4552 MPa within few seconds, as reported in [15]. It should be noted that without cooling, the chip will evaporate, and nothing will be left to talk about. Therefore, we need an efficient cooling mechanism to ensure that both temperature and stress do not exceed the safety limit. Embedded liquid-cooling channels may present an efficient way of cooling if the appropriate inlet velocity is used to optimize the cooling process.

### A. Low Inlet velocity

First, we set the inlet velocity into the embedded channels, shown in Fig. 2, to a low value of 10 mm/s to cool the chip. The corresponding temperature and stress are simulated. Figs. (3), (4), and (5) show the map of the temperature and the corresponding stress for water, R22, and LN cooling, respectively. The figure shows that the temperature of the chip is ranging from about 293 K to 420 K or 20 ˚C to 147 ˚C. When R22 is used instead, the temperature ranged from 240K-380K or -33˚C to 107˚C; see Fig. 4 (a). Although R22 reduced the temperature compared with cooling with water, the hot-spot temperatures are still unacceptable for both cooling techniques. To reduce the temperature to an acceptable value, we have two choices: either to use a different coolant or to increase the inlet velocity. We started with using a different coolant, LN. For LN, temperature is reduced ranging from 77 K to 220 K or -196 ˚C to -73 ˚C; see Fig. 5 (a).

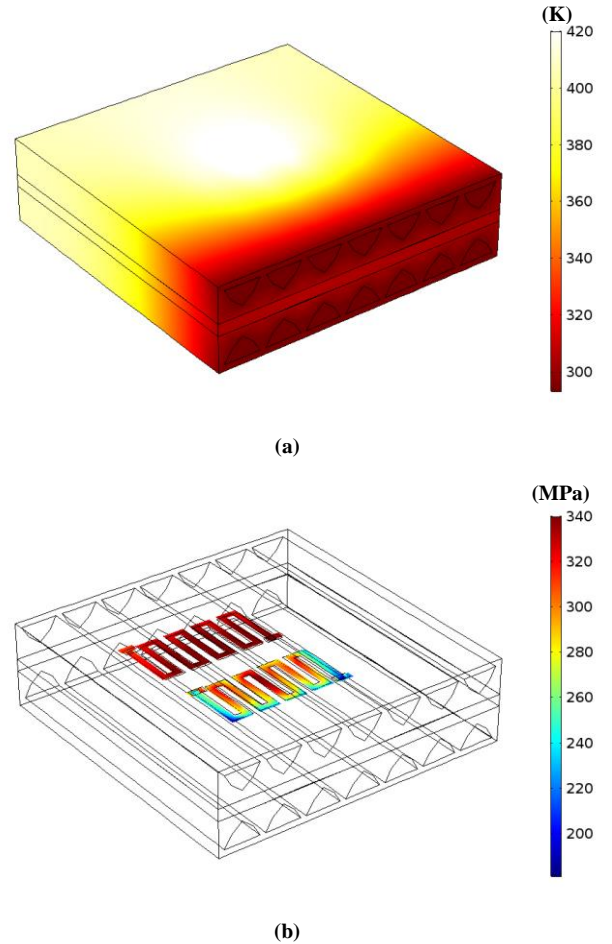

Fig.3: Water cooling with 10mm/s Pumping Velocity **(a)** Temperature and **(b)** Thermal Stress Map across Cu and Tungsten.

As can be seen from Figs 3(a), 4(a), and 5(a), the chip temperature at the inlet is the lowest and is equal to the temperature of the coolant liquid itself. As the liquid flows inside the block, it cools the IC by absorbing the heat from the chip. At the same time, the liquid temperature increases and its efficiency in cooling decreases. Hence the chip cooling rate slows down, and the temperature grows above the inlet temperature. The chip temperature increases as we go along the channels, and it may reach unacceptably high values, forming hotspots. These high temperatures are unacceptable because of their value and non-uniformity. The high temperature values may cause failure; and the non-uniformity may cause performance degradation of the chip.

responsible for the disappearance of overshooting. To the contrary, in [13] conduction takes-place through a low thermal conductivity $SiO_2$ layer of 1.5 $Wm^{-1}K^{-1}$. Therefore, using materials with higher thermal conductivities will result in increasing heat dissipation.

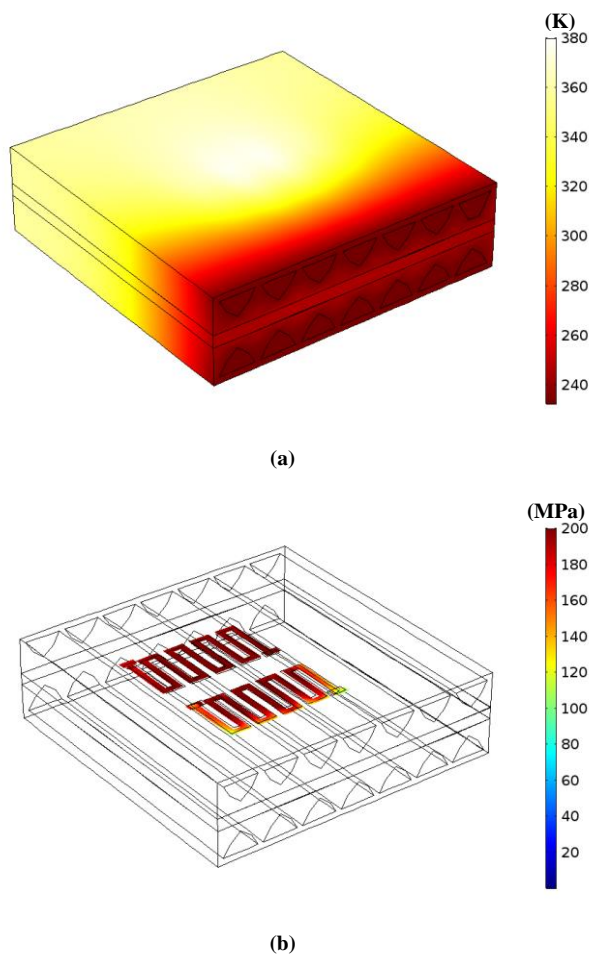

Fig.4: R22 cooling with 10mm/s Pumping Velocity **(a)** Temperature and **(b)** Thermal Stress Map across Cu and Tungsten.

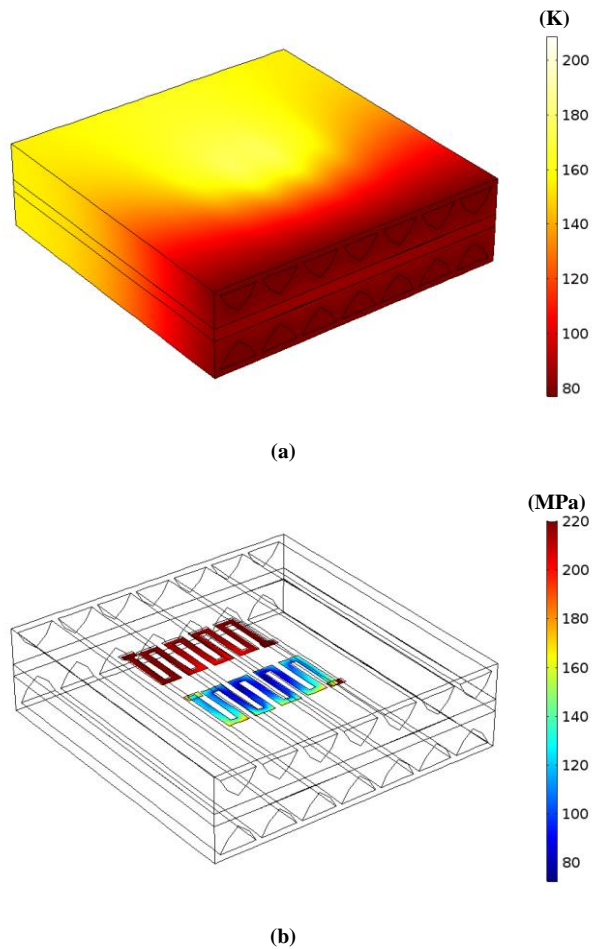

Fig.5: LN cooling with 10mm/s Pumping Velocity **(a)** Temperature and **(b)** Thermal Stress Map across Cu and Tungsten.

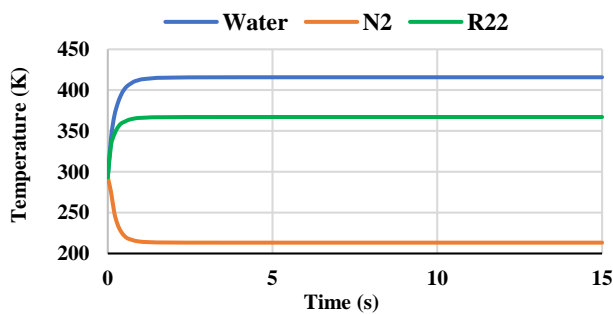

Fig.6: Temperature at 10mm/s Inlet Velocity at Cu Wire.

To probe the temperature variation more closely, the temperature of the Copper wire was observed over 15 seconds. Fig.6 shows that the temperature rises as an exponential function, then saturates within 2s. The temperature saturates to 415K, 367K and 213K for water, R22, and LN, respectively. These values are close to the maximum temperature observed in Figs 3(a), 4(a), and 5(a).

One observation from the figure is that there is no overshooting, as the case of [13]. The disappearance of temperature overshoot is believed to be due to the higher rate of cooling. Higher rate of cooling can be achieved either by increasing heat convection (by increasing liquid velocity), increasing heat conduction, or increasing radiation. For similar liquid velocities and outside atmospheres, it appears that the high thermal conductivity of Si (130 $Wm^{-1}K^{-1}$) is

Temperature non-uniformity should raise serious concern, since it is expected to result in the non-uniformity of the material properties across the chip and this shall lead to the non-uniformity of the characteristics of identical devices across the chip, and this leads to an added level of degradation. The problem is not only related to the electrical or optical performance of the devices. The problem extends to the physical structure of the chip. Among these problems, the creation of high thermal stress at the hotspots is one of the biggest concerns, which may lead to fatigue, fracture, or bulging of the films.

Stress on thin film metals is investigated. Figs. 3(b), 4(b), and 5(b) show maps for the von Mises stress on the hotspot (tungsten heater) and the copper wire, when chip is cooled using H$_2$O, R22, and LN. For all coolants, it is noticed that the stress on the Cu wire is higher that on tungsten. This is due to two factors: the first is the thermal expansion for copper is higher than tungsten, 16.5x10$^{-6}$/K compared with 4.5x10$^{-6}$/K. The second, is the fact that, in general, tungsten temperature is lower because it is closer to the inlet.

For water cooling, the highest temperature difference of the Cu is 122 ˚C (415 K-293 K). For R22 cooling, that difference is 74 ˚C, and for LN the difference is -80 ˚C. Therefore, the von Mises stress on Cu is 337 MPa for water, 200MPa for R22, and 211MPa for LN, see Fig.7. For water cooling, the resulting maximum stress is beyond both the tensile yield strength (33.3 MPa) and the ultimate tensile strength (of 210 MPa) for pure Copper [16]. For electroplated copper, with Poisson's ratio of 0.35 and thickness less than 0.885µm, the yield strength reaches 300MPa [17]. This higher yield strength is not high enough to prevent failure or fracture of the copper [18]. For cooling with R22 or LN, pure copper will deform or fracture under the resulting stress, but electroplated Cu can withstand the stress without a problem.

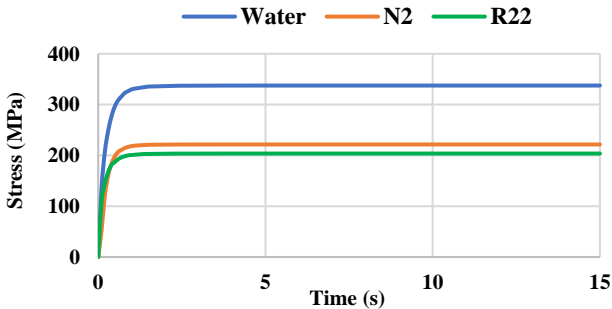

Fig.7: Stress at 10mm/s Inlet Velocity at Cu Wire.

The maps for the local stress are shown in Figs 3(b), 4(b) and 5(b), for water, R22, and LN cooling, respectively. From Fig. 3(b) and 4(b), it can be realized that the stress for the tungsten increases as we move far from the inlet, while it decreases on the Cu wire. This is because, for tungsten, the side closer to the inlet is exposed to colder liquid and has lower temperature. While, for Cu, the side closer to the inlet is closer to the heater and has higher temperature. This remains valid, as long as the stress is tensile, and the temperature is above the room temperature. For LN cooling, Fig. 5(b), the stress behavior is the opposite, even though the temperature behavior is the same as the other two cases with R22 and water. This can be explained by the fact that LN results in a compressive stress which increases as the temperature decreases. Hence, it is expected that the colder side will have higher stress and the warmer side will have lower stress, as long as the temperature is lower than the room temperature.

For tungsten heater the stress is tensile when coolant is H$_2$O or R22. For water cooling, the highest stress is found to range from 200 MPa at the sides to about 320 MPa at the middle. For cooling using R22, the stress ranges from about 100 MPa to 220 MPa. For LN, the stress is compressive ranging from about 80 MPa to 156 MPa. In all cases, the stress on tungsten is much less than the yield strength of 750 MPa and the ultimate tensile strength of 980 MPa [19]. In fact, tungsten yield can reach 1670 MPa and the ultimate strength to 3900 MPa [20].

The stress observed on the Aluminum pads are tensile for water and R22 reaching 318 MPa and 206 MPa, respectively. For LN, the stress is compressive reaching 148 MPa. Aluminum 6061 has a tensile yield strength of 267 MPa and maximum tensile strength of 310 MPa [21]. The results indicate that, except for water cooling, Al 6061 will not deform under cooling with R22 and LN. Stress on Al is determined by its thermal expansion (23x10$^{-6}$ K$^{-1}$) and temperature difference. Since, Al pads are located at the sides of the heater, hence they have lower temperatures.

*B. Inlet velocity of 100 mm/s*

The inlet velocity was increased to 100 mm/s. The temperature and von Mises stress are mapped in Figs. 8, 9 and 10 for water, R22 and liquid N$_2$ cooling, respectively.

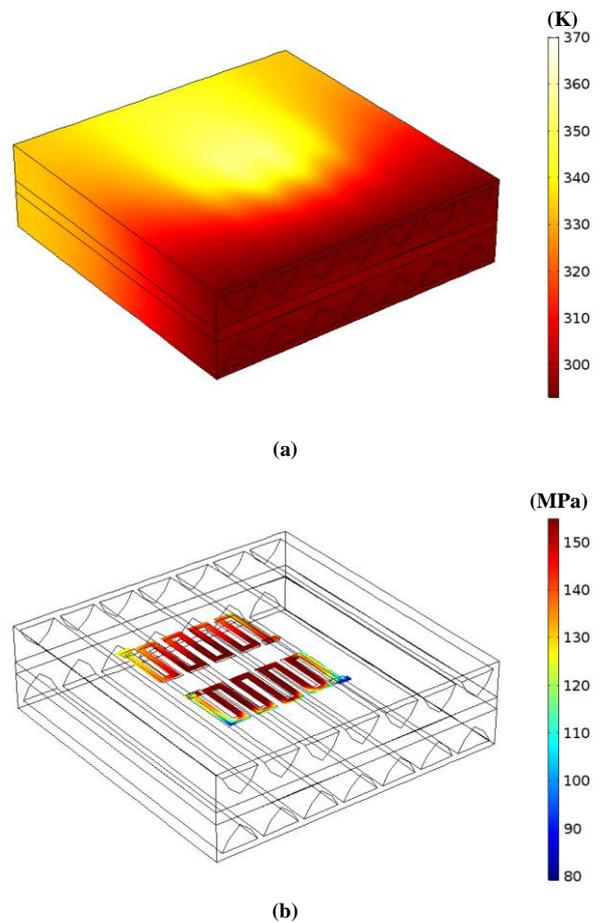

Fig.8: Water cooling with 100 mm/s Pumping Velocity **(a)** Surface Temperature Map and **(b)** Thermal Stress Map across Cu and Tungsten.

Fig. 11 shows that, after 1 s, the temperature values are reduced to 345K, 321K and 164K for water, R22 and LN, respectively. This results in a temperature difference from room temperature of 52K, 28K, and -129K, for water, R22, and LN, respectively. Proportional to this temperature difference, Fig. 12 shows that the stress on the copper wire reach saturation at 144MPa, 78MPa and 355MPa respectively for water, R22 and LN. Bulk Cu will yield and deform at these stresses, since the yield strength is 33.3 MPa. If electroplated Cu is used, metal will not deform for water cooling although LN cooling would result in deformation of Cu.

Our simulation shows that, at the tungsten hotspot, the temperature saturates at 364 K for water, 336 K for R22, and 179 K for LN. The stress observed at the hotspot are 150 MPa, 94 MPa and 222 MPa for water, R22 and liquid nitrogen, respectively. For tungsten hotspot, the stress values are far below the tensile and compressive yield stress. The stress on the aluminum pads are 158 MPa, 104 MPa, and 242 MPa for water, R22, and LN. Since the tensile yield strength of Al 6061 is 267 MPa, Al pads will be resistant to deformation at these stresses [21].

It is quite normal for a processor to have an idle temperature around 318K to 323K (25 ˚C to 30 ˚C) and full load temperature (50 ˚C to 60 ˚C) [22]. Therefore, all cooling techniques at 100 mm/s velocity can satisfy the temperature requirement for this case. However, electroplated Cu must be used to ensure higher yield strength or other metal structures with higher yield need to be used.

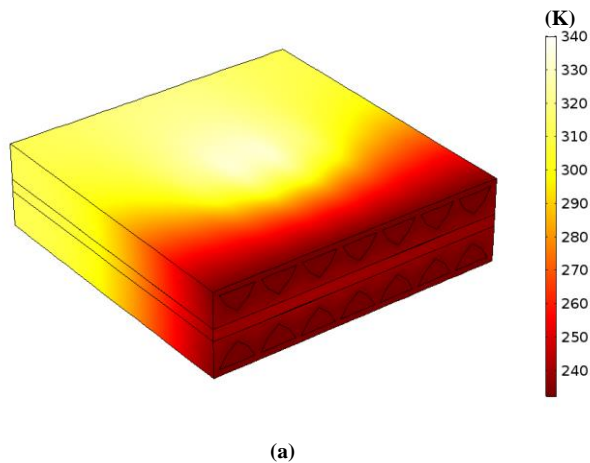

(a)

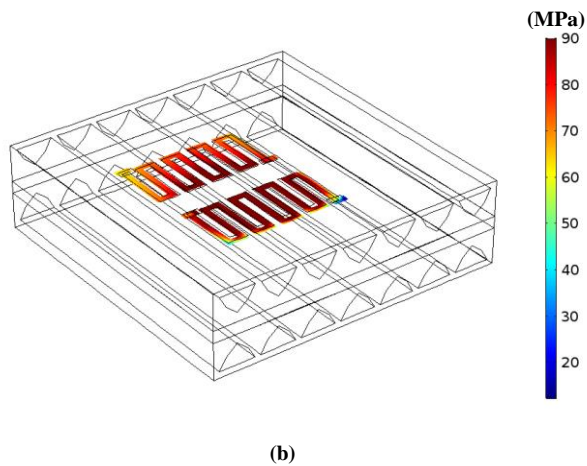

(b)

Fig.9: R22 cooling with 100mm/s Pumping Velocity **(a)** Surface Temperature Map and **(b)** Thermal Stress Map across Cu and Tungsten.

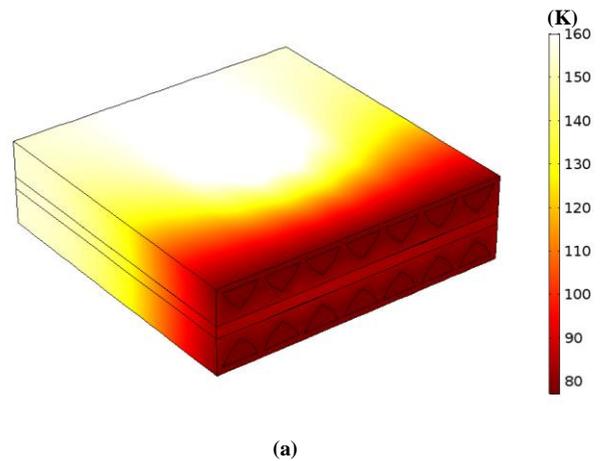

(a)

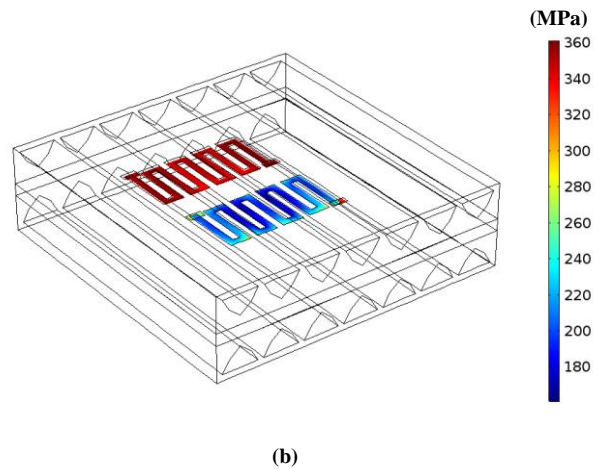

(b)

Fig.10: LN cooling with 100mm/s Pumping Velocity **(a)** Surface Temperature Map and **(b)** Thermal Stress Map across Cu and Tungsten.

## V. CONCLUSION

The study investigates the temperature and stress on the different metal wires and pads, when a 20 W hotspot is cooled using embedded channels. The study considered three different coolants, water, R22, and LN with admission velocities of 10 mm/s and 100 mm/s. The study revealed important facts. First, water may not be able to bring temperature to an acceptable level to avoid Cu and Al deformation, especially at lower fluid velocity of 10 mm/s. Second, bulk copper cannot withstand the stress using any coolant, due to the low yield strength of 33.3 MPa. Third, if electroplated Cu is used, the yield strength increases to 300 MPa, and the metal can withstand the stress resulting from cooling using R22 and LN at 10mm/s velocity. Fourth, if the velocity increases to 100 mm/s, metal can withstand stress when water and R22 is used, but LN cooling causes unacceptable stress. In other word, R22 is the most suitable coolant for cooling 20 W hotspot. Since Al 6061 has a yield of 267 MPa, what is said about electroplated Cu can be said about Al alloy. For tungsten, with a yield strength of a minimum 750 MPa, no deformation will take-place under any of the cooling mechanisms.

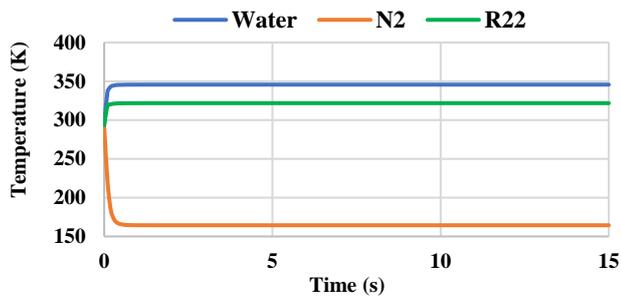

Fig.11: Temperature at 100mm/s Inlet Velocity Cu Wire.

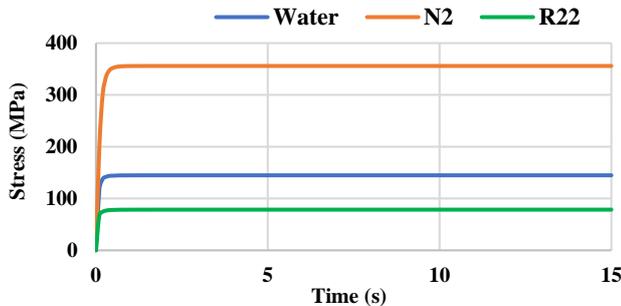

Fig.12: Stress at 100mm/s Inlet Velocity Cu Wire.

The study also shows that admitting the fluid from one side results in a large temperature and stress gradient that can reach 140 ˚C and 100 MPa, respectively. This creates local hot-spots and high-stress spots. These spots may not damage the IC, but it may make the performance unacceptable. The study shows the need for redesigning the embedded channel system in order to optimize the cooling process and eliminate both hot-spots and high-stress spots.